\documentclass[preprint,showpacs,showkeys,preprintnumbers,amsmath,amssymb,
eqsecnum,prd]{revtex4}
\usepackage{epsfig}
\usepackage{dcolumn}% Align table columns on decimal point
\usepackage{bm}% bold math

{\hspace*{\fill}{\protect\small 
{\bf Bijan Saha and Todor Boyadjiev}}
\hspace*{\fill}
}
{\hspace*{\fill}
{\protect\small {\bf Interacting spinor and scalar fields in a Bianchi 
type-I Universe ...}}
\hspace*{\fill}                        
}
\newcommand {\ve}{\varepsilon}
\newcommand {\pr}{\partial}
\newcommand {\cG}{\cal G}
\newcommand {\cD}{\cal D}
\newcommand {\bg}{\bar \gamma}
\newcommand {\G}{\Gamma}
\newcommand {\bp}{\bar \psi}

\newcommand {\vf}{\varphi}

\begin{document}
%\condensed
\title{Interacting spinor and scalar fields in a Bianchi type-I Universe: 
Oscillatory solutions}
\author{Bijan Saha and Todor Boyadjiev} 
\affiliation{Laboratory of Information Technologies\\ 
Joint Institute for Nuclear Research, Dubna\\ 
141980 Dubna, Moscow region, Russia} 
\email{saha@thsun1.jinr.ru, todorlb@jinr.ru}

\begin{abstract}
Self-consistent system of spinor, scalar and BI gravitational 
fields is considered. Exact solutions to the field equations in terms
of volume scale of the BI metric are obtained. Einstein field equations
in account of the cosmological constant $\Lambda$ and perfect fluid
are studied. Oscillatory mode of expansion of the universe
is obtained. It is shown that for the interaction
term being a power function of the invariants of bilinear spinor forms
and $\Lambda > 0$ and given other parameters, e.g., 
coupling constant, spinor mass etc., there exists a finite range of 
integration constant which generates oscillatory mode of evolution.  
\end{abstract}
\vskip 5mm
\noindent
\vskip 5mm
\noindent

\maketitle
%%%%%%%%%%%%%%%%%%%%%%%%%%%%%%%%%%%%%%%%%%%%%%%%%%%%%%%%%%%%%%%%%%%%%%%%%%%%

\section{Introduction}

The discovery of the cosmic microwave radiation has stimulated a growing
interest in anisotropic, general-relativistic cosmological models of the
universe. The choice of anisotropic cosmological models in the system of
Einstein field equation enable us to study the early day universe, which
had an anisotropic phase that approaches an isotropic one~\cite{misner}.
BI universe is the simplest model of an anisotropic universe and eventually
evolves into a Freidmann-Robertson-Walker (FRW) universe~\cite{jacobs}, 
if filled with matter obeying  $p\,=\,\zeta\,\ve, \quad 
\zeta < 1$, where $\ve$ and $p$ are the energy and pressure of the
material field, respectively. Since the present-day universe is 
surprisingly isotropic, this feature of the BI Universe makes it a prime 
candidate for studying the possible effects of an anisotropy in the early 
universe on present-day observations.

In this paper we study the self-consistent system of spinor, scalar and
BI gravitational fields in presence of perfect fluid. Solutions of Einstein 
equations coupled to a spinor and a scalar fields in BI spaces have
been extensively studied by Saha and Shikin 
~\cite{sahagrg,sahajmp,sahaprd,sahal}. In those papers, the field equations
were solved qualitatively. In this report, we consider some key equations
occurred in those papers. Initial value problem has been posed for those 
equations and solved numerically. Application of numerical methods enables us
to view this problem from totally different angle and gives rise to some 
interesting results previously unknown.

%%%%%%%%%%%%%%%%%%%%%%%%%%%%%%%%%%%%%%%%%%%%%%%%%%%%%%%%%%%%%%%%%%%%%%%%%%%%%

\section{Fundamental Equations and general solutions}

The action of the nonlinear spinor, scalar and gravitational fields
can be written as
\begin{equation}
{\cal S}(g; \psi, \bp, \vf) = \int\, (R + L) \sqrt{-g} d\Omega,
\label{action}
\end{equation}
where $R$ is the Ricci scalar and $L$ is the spinor and scalar field 
Lagrangian density chosen in the form\cite{sahagrg} 
\begin{equation} 
L= \frac{i}{2} 
\biggl[\bp \gamma^{\mu} \nabla_{\mu} \psi- \nabla_{\mu} \bar 
\psi \gamma^{\mu} \psi \biggr] - m\bp \psi +
\frac{1}{2} \vf_{,\alpha}\vf^{,\alpha} (1 + \lambda F).
\label{lag} 
\end{equation} 
Here $\lambda$ is the coupling constant and $F$ is some 
arbitrary functions of invariants generated from the real bilinear 
forms of a spinor field. We choose $F$ to be the 
function of $I =S^2 = (\bp \psi)^2$ and 
$J = P^2 = (i \bp \gamma^5 \psi)^2$, i.e., $F = F(I, J)$, 
that describes the nonlinearity in the most general of its 
form~\cite{sahaprd}. As one sees, for $\lambda = 0$ we have the
system with minimal coupling. 

The gravitational field in our
case is given by a Bianchi type I (BI) metric in the form 
\begin{equation} 
ds^2 = a_0^2 (dx^0)^2 - a_1^2 (dx^1)^2 - a_2^2 (dx^2)^2 - a_3^2 (dx^3)^2, 
\label{BI1}
\end{equation}
with $a_0 = 1$, $x^0 = c t$ and $c = 1$. The metric functions
$a_i$ $(i=1,2,3)$ are the functions of time $t$ only.

Variation of \eqref{action} with respect to spinor field $\psi\,(\bp)$
gives nonlinear spinor field equations
\begin{subequations}
\label{speq}
\begin{eqnarray}
i\gamma^\mu \nabla_\mu \psi - m \psi + {\cD} \psi + 
{\cG} i \gamma^5 \psi &=&0, \label{speq1} \\
i \nabla_\mu \bp \gamma^\mu +  m \bp - {\cD} \bp - 
{\cG} i \bp \gamma^5 &=& 0, \label{speq2}
\end{eqnarray}
\end{subequations}
where we denote
$$ {\cD} =  \lambda S \vf_{,\alpha}\vf^{,\alpha} {\pr F}/{\pr I}, 
\quad
{\cG} =  \lambda P \vf_{,\alpha}\vf^{,\alpha} {\pr F}/{\pr J},$$
whereas, variation of \eqref{action} with respect to scalar field
yields the following scalar field equation
\begin{equation}
\frac{1}{\sqrt{-g}} \frac{\pr}{\pr x^\nu} \Bigl(\sqrt{-g} g^{\nu\mu}
(1 + \lambda F) \vf_{,\mu}\Bigr) 
= 0. \label{scfe}
\end{equation}

Varying \eqref{action} with respect to metric tensor $g_{\mu\nu}$ 
one finds the gravitational field equation which in account of cosmological 
constant $\Lambda$ has the form 
\begin{subequations}
\label{BID}
\begin{eqnarray}
\frac{\ddot a_2}{a_2} +\frac{\ddot a_3}{a_3} + \frac{\dot a_2}{a_2}\frac{\dot 
a_3}{a_3}&=&  \kappa T_{1}^{1} -\Lambda,\label{11}\\
\frac{\ddot a_3}{a_3} +\frac{\ddot a_1}{a_1} + \frac{\dot a_3}{a_3}\frac{\dot 
a_1}{a_1}&=&  \kappa T_{2}^{2} - \Lambda,\label{22}\\
\frac{\ddot a_1}{a_1} +\frac{\ddot a_2}{a_2} + \frac{\dot a_1}{a_1}\frac{\dot 
a_2}{a_2}&=&  \kappa T_{3}^{3} - \Lambda,\label{33}\\
\frac{\dot a_1}{a_1}\frac{\dot a_2}{a_2} +\frac{\dot a_2}{a_2}\frac{\dot 
a_3}{a_3}+\frac{\dot a_3}{a_3}\frac{\dot a_1}{a_1}&=&  \kappa T_{0}^{0} - 
\Lambda.
\label{00}
\end{eqnarray}
\end{subequations}
Here $\kappa$ is the Einstein gravitational constant and over-dot means 
differentiation with respect to $t$. The energy-momentum tensor of the 
material field is given by
\begin{eqnarray}
T_{\mu}^{\rho} &=& \frac{i}{4} g^{\rho\nu} \biggl(\bp \gamma_\mu 
\nabla_\nu \psi + \bp \gamma_\nu \nabla_\mu \psi - \nabla_\mu \bar 
\psi \gamma_\nu \psi - \nabla_\nu \bp \gamma_\mu \psi \biggr)  \label{tem}\\
& & + (1 - \lambda F)  \vf_{,\mu}\vf^{,\rho} - \delta_{\mu}^{\rho} L
+ T_{{\rm m}\,\mu}^{\,\,\,\nu}. \nonumber
\end{eqnarray}
Here $T_{\mu\,(m)}^{\nu} = (\ve,\,-p,\,-p,\,-p)$
is the energy-momentum tensor of a perfect fluid. 
Energy $\ve$ is related to the pressure $p$ by the equation 
of state $p\,=\,\zeta\,\ve$.  Here $\zeta$ varies between the
interval $0\,\le\, \zeta\,\le\,1$, whereas $\zeta\,=\,0$ describes
the dust Universe, $\zeta\,=\,\frac{1}{3}$ presents radiation Universe,
$\frac{1}{3}\,<\,\zeta\,<\,1$ ascribes hard Universe and $\zeta\,=\,1$
corresponds to the stiff matter. The Dirac matrices $\gamma_\mu(x)$
of curve space-time are connected with those of Mincowski space as
\begin{equation}
\gamma^\mu = \bg^\mu/a_\mu, \quad \gamma_\mu = \bg a_\mu, \quad
\mu = 0,1,2,3. 
\end{equation}

The explicit form of the covariant derivative
of spinor is ~\cite{brill}
\begin{eqnarray} 
\nabla_\mu \psi = \partial_\mu \psi - \G_\mu \psi, \quad
\nabla_\mu \bp = \partial_\mu \bp + \bp \G_\mu, 
\quad \mu = 0,1,2,3,
\label{cvd}
\end{eqnarray} 
where $\G_\mu(x)$ are spinor affine connection matrices.  
For the metric \eqref{BI1} one has the following components
of the affine spinor connections 
\begin{equation}  
\G_\mu = (1/2){\dot a_\mu} \bg^\mu \bg^0.
\label{safc}
\end{equation}
 
We study the space-independent solutions to the spinor 
and scalar field Eqns. \eqref{speq} and \eqref{scfe} so that 
$\psi=\psi(t)$ and $\vf = \vf(t)$.
defining
\begin{equation}
\tau = a_0 a_1 a_2 a_3 = \sqrt{-g}
\label{taudef}
\end{equation}
from \eqref{scfe} for the scalar field  we have
\begin{equation}
\vf = C \int [\tau (1 + \lambda F)]^{-1} dt.
\label{sfsol}
\end{equation}

Setting $V_j(t) = \sqrt{\tau} \psi_j(t), \quad j=1,2,3,4,$ in view
of \eqref{cvd} and \eqref{safc} 
from \eqref{speq1} one deduces the following system of equations:  
\begin{subequations}
\label{V}
\begin{eqnarray} 
\dot{V}_{1} + i (m - {\cD}) V_{1} - {\cG} V_{3} &=& 0, \\
\dot{V}_{2} + i (m - {\cD}) V_{2} - {\cG} V_{4} &=& 0, \\
\dot{V}_{3} - i (m - {\cD}) V_{3} + {\cG} V_{1} &=& 0, \\
\dot{V}_{4} - i (m - {\cD}) V_{4} + {\cG} V_{2} &=& 0. 
\end{eqnarray} 
\end{subequations}

>From \eqref{speq1} we also write the equations for the bilinear spinor 
forms $S,\quad P$ and $A^0 = \bp \bg^5 \bg^0 \psi$
\begin{subequations}
\begin{eqnarray}
{\dot S_0} - 2 {\cG}\, A_0^0 &=& 0, \label{S0}\\
{\dot P_0} - 2 (m - {\cD})\, A_0^0 &=& 0, \label{P0}\\
{\dot A_0^0} + 2 (m - {\cD})\, P_0 + 2 {\cG} S_0 &=& 0, \label{A0} 
\end{eqnarray}
\end{subequations}
where $Q_0 = \tau Q$, leading to the relation
$S^2 + P^2 + (A^0)^2 =  C^2/ \tau^2, \qquad C^2 = {\rm const.}$
As one sees, for $F=F(I)$ \eqref{S0} gives $S = C_0/\tau$,
whereas for the massless spinor field with $F=F(J)$ \eqref{P0}
yields $P=D_0/\tau$. In view of it for $F=F(I)$ we obtain the
following expression for the components of spinor field
\begin{eqnarray} 
\psi_1(t) &=& C_1 \tau^{-1/2} e^{-i\beta}, \quad
\psi_2(t) = C_2 \tau^{-1/2} e^{-i\beta},  \nonumber\\
\label{spef}\\
\psi_3(t) &=& C_3 \tau^{-1/2} e^{i\beta}, \quad
\psi_4(t) = C_4 \tau^{-1/2} e^{i\beta},
\nonumber
\end{eqnarray} 
with $C_i$ being the integration constants and
are related to $C_0$ as 
$C_0 = C_{1}^{2} + C_{2}^{2} - C_{3}^{2} - C_{4}^{2}.$ Here
$\beta = \int(m - {\cD}) dt$. In case of $F=F(J)$ for the massless 
spinor field we get 
\begin{eqnarray}
\psi_1 &=& \tau^{-1/2} \bigl(D_1 e^{i \sigma} + 
iD_3 e^{-i\sigma}\bigr), \quad
\psi_2 = \tau^{-1/2} \bigl(D_2 e^{i \sigma} + 
iD_4 e^{-i\sigma}\bigr), \nonumber \\
\label{J}\\
\psi_3 &=& \tau^{-1/2} \bigl(iD_1 e^{i \sigma} + 
D_3 e^{-i \sigma}\bigr),\quad
\psi_4 = \tau^{-1/2} \bigl(iD_2 e^{i \sigma} + 
D_4 e^{-i\sigma}\bigr). \nonumber
\end{eqnarray} 
The integration constants $D_i$
are connected to $D_0$ by
$D_0=2\,(D_{1}^{2} + D_{2}^{2} - D_{3}^{2} -D_{4}^{2}).$
Here we set $\sigma = \int {\cG} dt$. 

Once the spinor functions are known explicitly, one can write the components 
of spinor current
$j^\mu = \bp \gamma^\mu \psi$,
the charge density of spinor field 
$\varrho = (j_0\cdot j^0)^{1/2}$,
the total charge of spinor field 
$Q = \int\limits_{-\infty}^{\infty} \varrho \sqrt{-^3 g} dx dy dz,$
the components of spin tensor
$
S^{\mu\nu,\epsilon} = \frac{1}{4}\bp \bigl\{\gamma^\epsilon
\sigma^{\mu\nu}+\sigma^{\mu\nu}\gamma^\epsilon\bigr\} \psi$
and other physical quantities.

Let us now solve the Einstein equations. In doing so we first write the 
expressions for the components of the energy-momentum tensor explicitly:
\begin{eqnarray}
\label{total}
T_{0}^{0} &=& mS + C^2/2\tau^2 (1+\lambda F) + \ve, \nonumber\\
\\
T_{1}^{1} &=& T_{2}^{2} = T_{3}^{3} =
{\cD} S + {\cG} P - C^2/2\tau^2 (1+\lambda F) 
- p.\nonumber 
\end{eqnarray}
In account of \eqref{total} from  \eqref{11}, \eqref{22}, \eqref{33}
we find the metric functions~\cite{sahaprd}
\begin{equation} 
a_i(t) = D_i \tau^{1/3}\mbox{exp}\biggl[ X_i 
 \int\,[\tau (t)]^{-1} dt \biggr],\quad i=1,2,3, \label{a}
\end{equation}
with the constants of integration $D_i$ and $X_i$ obeying
$$\prod_{i=1}^{3} D_i = 1, \quad \sum_{i=1}^{3} X_i = 0.$$

As one sees from \eqref{a} for $\tau \sim t^n$
with $n > 1$ the exponent tends to unity at large $t$, and the 
anisotropic model becomes isotropic one. 
Let us also write the invariants of gravitational field. They are
the Ricci scalar $I_1 = R \approx 1/\tau$,
$I_2 = R_{\mu\nu}R^{\mu\nu} \equiv R_{\mu}^{\nu} R_{\nu}^{\mu}
\approx 1/\tau^3$ and the Kretschmann scalar
$I_3 = R_{\alpha\beta\mu\nu}R^{\alpha\beta\mu\nu} \approx 1/\tau^6$.
As we see, the space-time becomes singular at a point where $\tau = 0$,
as well as the scalar and spinor fields. Thus we see, all the functions
in question are expressed via $\tau$. In what follows, we write the
equation for $\tau$ and study it in details.  

Summation of Einstein Eqns. \eqref{11}, \eqref{22}, \eqref{33} and 
\eqref{00} multiplied by 3 gives
\begin{equation}
\frac{\ddot 
\tau}{\tau}= \frac{3}{2}\kappa \Bigl(mS + {\cD} S + {\cG} P + \ve -p
\Bigr) - 3 \Lambda. 
\label{dtau1}
\end{equation} 

From energy-momentum conservation law $T_{\mu;\nu}^{\nu} = 0$,
in account of the equation of state $p = \zeta \ve$, we obtain 
\begin{equation}
\ve = {\ve_0}/{\tau^{1+\zeta}},\quad 
p = {\zeta \ve_0}/{\tau^{1+\zeta}}.
\label{vep}
\end{equation}
In our consideration  of $F$ as a function of $I$, $J$ or $I\pm J$
we get these arguments, as well as ${\cD}$ and ${\cG}$ as functions of
$\tau$. Hence the right-hand-side of \eqref{dtau1} is a function
of $\tau$ only. In what follows we consider the case with $F=F(I)$.
 Recalling the definition of ${\cD}$,
in view of \eqref{sfsol} and \eqref{vep} the Eqn. \eqref{dtau1} can be 
written as
\begin{equation} \label{dtau2}
    {\ddot \tau} = \mathcal F(\tau, p)\,,
\end{equation}
where we denote
\begin{equation}\label{pot}
   \mathcal F \equiv \frac{3}{2} \kappa \Bigl(m C_0 + \lambda C_0^2 C^2 F_I
(\tau) /\tau^3 (1 + \lambda F(\tau))^2 + \ve_0(1-\zeta)/\tau^{\zeta}\Bigr) - 
3 \Lambda \tau\,,
\end{equation}
and 
$p \equiv \{\kappa, \lambda, m, C_0, C, \varepsilon_0, \zeta, \Lambda \}$ is 
the set of the parameters. Here we take into account that $ S = C_0/\tau$. 
From mechanical point of view the Eqn. \eqref{dtau2} can be interpreted 
as an equation of motion of a single particle with unit mass under the 
force $\mathcal F(\tau,p)$. Then the following first integral exists 
\cite{ll_76}
\begin{equation} \label{1stint}
    \dot \tau = \sqrt{E - \mathcal U(\tau,p)}\,.
\end{equation}
Here $E$ is the integration constant and 
$$\mathcal{U} \equiv - \frac{3}{2} \Bigl[\kappa \Bigl(m C_0 \tau -
C^2/2(1 + \lambda F) 
+ \ve_0 \tau^{-\zeta}\Bigr) - \Lambda \tau^2\Bigr]\,,$$
is the potential of the force $\mathcal F$. We note that the radical 
expression must be non-negative. The zeroes of this expression, which 
depend on all the problem parameters $p$ define the boundaries of the 
possible rates of changes of $\tau(t)$. Note that setting $m = 0$ in 
\eqref{pot} we come to the case of massless spinor field with $F=F(J)$ 
or $F=F(I \pm J)$.

We formulate the initial value problem for the Eqn. \eqref{dtau2} with 
initial condition
$$\tau(0) = \tau_0 > 0\,,$$
which we solve numerically. 

Since $\tau$ is the volume-scale, it cannot be negative for every $t\geq 0$. 
On the other-hand, BI space-time models a non-static universe, i.e., the 
derivative ${\dot \tau(t)}$ should be nontrivial at the initial moment 
$t = 0$. This leads to to fact that for fixed $p$, the constant $E$ and
the initial value of $\tau$ are inter-related in the sense that for a given 
$E$ the value of $\tau_0$ should belong to some interval.

In what follows, we numerically solve \eqref{dtau2} for some concrete form
of $F$. 

Let us choose $F$ as a power function of $S$, namely,
$F = S^n$. In this case setting $C_0 = 1$ and $C = 1$ we obtain
\begin{equation}
 \mathcal F = \frac{3\kappa}{2}\Bigl(m + \frac{\lambda\, n\, \tau^{n-1}}{2\, 
(\lambda  + \tau^n)^2} + \ve_0 \frac{(1-\zeta)}{\tau^\zeta} \Bigr) - 3 
\Lambda\, \tau, \label{nueq}
\end{equation}
with the potential
\begin{equation}
  \mathcal{U} = - \frac{3}{2} \left\{\kappa \left [m\, \tau - \frac{\lambda}
{2\,(\lambda + \tau^n) } \right] - \Lambda \tau^2 +\ve_0 \tau^{1-\zeta}
\right\}. \label{quads}
\end{equation}
Note that the nonnegativity of the radical in \eqref{1stint} in view of 
\eqref{quads} imposes restriction on $\tau$ from above in case of 
$\Lambda > 0$. It means that in case of $\Lambda > 0$ the value of 
$\tau$ runs between $0$ and some $\tau_{\rm max}$, where $\tau_{\rm max}$ 
is the maximum value of $\tau$ for the given value of $p$. 
This equation has been studied for different values of parameters $p$. 
Here we demonstrate the evolution of $\tau$ for different choice of 
$\tau_0$ for fixed ``energy'' $E$ and vise versa.

As the first example we consider massive spinor field with $m = 1$.
Other parameters are chosen in the following way: 
coupling constant $\lambda = 0.1$, power of nonlinearity $n = 4$, 
and cosmological constant $\Lambda = 1/3$. We also choose $\zeta = 0.5$
describing a hard universe.

\begin{figure}
    \begin{center}
        \epsfig{file=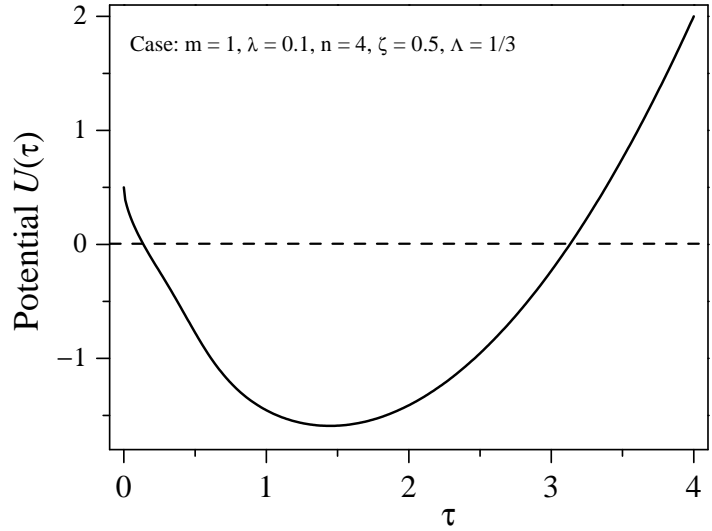,height=7cm} 
\caption{Perspective view of the potential $\mathcal U(\tau)$.}
 \label{pot1}
    \end{center}
\end{figure}

\begin{figure}
    \begin{center}
        \epsfig{file=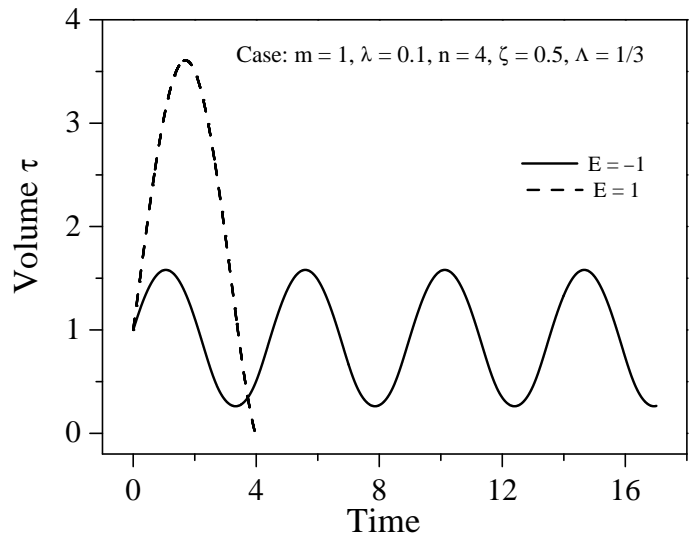,height=7cm} \caption{Perspective view of $\tau$ for different choice of $E$.}
 \label{tau1}
    \end{center}
\end{figure}

In Fig. \ref{pot1} we plot corresponding potential $\mathcal U(\tau)$ 
multiplied by the factor $2/3$. As is seen from Fig.~\ref{pot1} and 
Fig.~\ref{tau1}, choosing the integration constant $E$ we may obtain two 
different types of solutions. For $E > 0.5$ solutions are non-periodic, 
whereas for $E_{\rm min} < E \le 0.5$ the evolution of the universe is
oscillatory.

As a second example we consider the massless spinor field. Other parameters
of the problem are left unaltered with the exception of $\zeta$. Here we
choose $\zeta = 1$ describing stiff matter. It should be noted that
this particular choice of $\zeta$ gives rise to a local maximum. 
This results in two types of solutions for a single choice of $E$.

\begin{figure}
    \begin{center}
        \epsfig{file=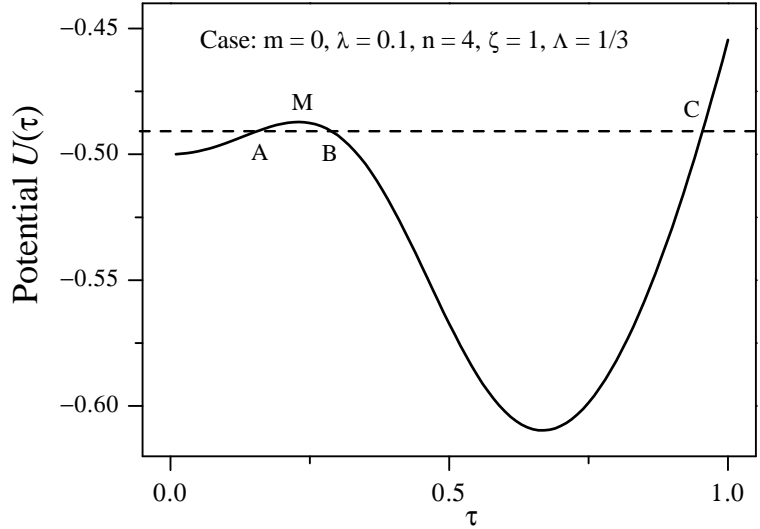,height=7cm} 
\caption{Perspective view of the potential $\mathcal U(\tau)$ with BI
universe being filled with stiff matter.}
 \label{pot2}
    \end{center}
\end{figure}

\begin{figure}
    \begin{center}
        \epsfig{file=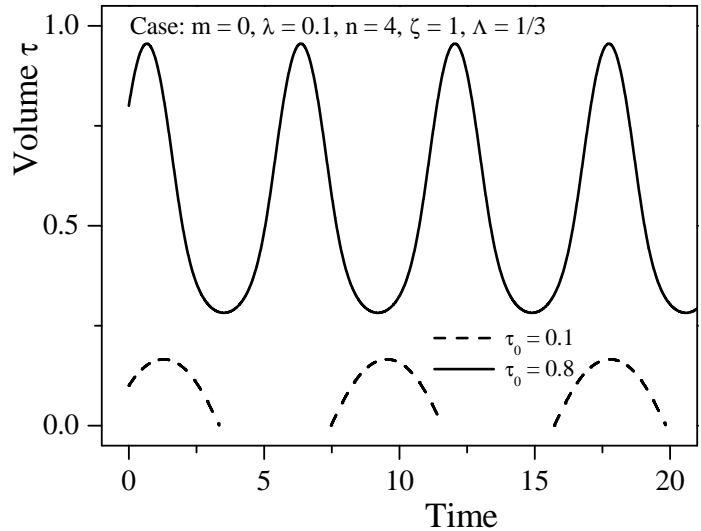,height=7cm} \caption{Perspective view of $\tau$ for different choice of $\tau_0$ with $E \in (-0.5,\, M)$.}
 \label{tau2}
    \end{center}
\end{figure}

As one sees from Fig.~\ref{pot2}, for $E > M$ there exists only non-periodic
solutions, whereas, for $E_{\rm min} < E < -0.5$ the solutions are always
oscillatory. For $E \in (-0.5,\, M)$ there exits two types of solutions
depending on the choice of $\tau_0$. In Fig.~\ref{tau2} we plot the 
evolution of $\tau$ for $E \in (-0.5,\, M)$. As is seen, for 
$\tau_0 \in (0,\, A)$ we have periodical solution, but due to the fact
that $\tau$ is non-negative, the physical solutions happen to be
semi-periodic. For $\tau_0 \in (B,\, C)$ we again have oscillatory mode
of the evolution of $\tau$. This two region is separated by a no-solution
zone $(A,\,B)$.

\section{Conclusions}

A self-consistent system of spinor, scalar and gravitation fields has been 
studied in presence of perfect fluid and cosmological term $\Lambda$. 
Oscillatory mode of evolution of the universe is obtained. It is shown that 
for the interaction term being a power function of the invariants of
bilinear spinor forms, the oscillatory solution is possible if 
$\Lambda$-term is positive. It is also shown that only for a finite range
the integration constant $E$ there exists oscillatory mode of evolution. It 
should be emphasized that a third type of solution is possible, if the BI
universe is filled with stiff matter.

\begin{acknowledgments}
We would like to thank Prof. E.P.~Zhidkov
for his kind attention to this work and helpful discussions. 

The work of T.L. B. was supported in part by Bulgarian Scientific Fund.
\end{acknowledgments}

%%%%%%%%%%%%%%%%%%%%%%%%%%%%%%%%%%%%%%%%%%%%%%%%%%%%%%%%%%%%%%%%%%%%%%%%%%%%%%%%

\end{document}